\documentclass[aps,amsmath,twocolumn,footinbib]{revtex4-2}
\usepackage{graphicx}
\usepackage{dcolumn} 
\usepackage{bm}
\usepackage[utf8]{inputenc}
\usepackage{amsmath}
\usepackage{hyperref}
\hypersetup{colorlinks,allcolors=black}
\usepackage{physics}
\usepackage{mymacros}
\usepackage{siunitx}
\usepackage[dvipsnames]{xcolor}
\usepackage[utf8]{inputenc}

\begin{document}
\title{Chiral Magic-Angle Twisted Bilayer Graphene in a Magnetic Field: \\
Landau Level Correspondence, Exact Wavefunctions and Fractional Chern Insulators}
\date{May 2021}
\author{Yarden Sheffer}
\email{yarden.sheffer@gmail.com}
\affiliation{Department of Condensed Matter Physics, Weizmann Institute of Science Rehovot 7610001, Israel}
\author{Ady Stern}
\email{adiel.stern@weizmann.ac.il}
\affiliation{Department of Condensed Matter Physics, Weizmann Institute of Science Rehovot 7610001, Israel}

\begin{abstract}
We show that the flat bands in the chiral model of magic-angle twisted bilayer graphene remain exactly flat in the presence of a perpendicular magnetic field. This is shown by an exact mapping between the model and the lowest Landau level wavefunctions at an effective magnetic field, in which the external field is either augmented or reduced by one flux quantum per unit cell. When the external field reaches one flux quantum per unit cell, the model exhibits a topological phase transition. These findings allow us to analyze a Jain-series of Fractional Chern Insulators states in the exactly flat band, and to point out an unconventional dependence of the energy gap on the magnetic field.  
\end{abstract}
\maketitle
Magic-angle twisted bilayer graphene (MATBG) \cite{PhysRevB.82.121407, Bistritzer2011a, Cao2018, Cao2018a,andrei2020graphene} has generated interest as a platform that combines two major components of modern condensed matter physics: strong interactions and topology. The first is a result of the flat bands occurring at charge neutrality \cite{PhysRevB.82.121407, Bistritzer2011a}. As for the second, the flat bands host "fragile" topology \cite{po2018origin,Po2018, Song2019, Ahn2019}, and can be decomposed into two Chern bands of opposite Chern numbers (per spin and valley) \cite{Bultinck2020,xie2021twisted}, which are connected by parity-time ($C_2\TT$) symmetry. As a result of this combination the phase diagram of MATBG was shown to host a wide range of phenomena including superconductivity \cite{Cao2018} (possibly unconventional \cite{Cao2018,Cao2021}), correlated insulators \cite{Cao2018a}, correlated Chern insulators \cite{Nuckolls2020a,Chen2020,Saito2021,Pierce2021}, and most recently fractional Chern insulators \cite{xie2021fractional}. Another point of interest in TBG (and in moir\'e materials in general) is that, as a result of the large lattice size, it is experimentally possible to apply a magnetic field of the order of a single flux quantum per unit cell \cite{Bistritzer2011, Dean2013, Saito2021, Herzog-Arbeitman2020}. This allows the experimental study phenomena such as the Hofstadter spectrum \cite{Hofstadter1976}, and even fractional states emerging on the Hofstadter subbands \cite{wang2015evidence,Spanton2018}. 

In the study of MATBG, a significant role was played by the so-called chiral model \cite{Tarnopolsky2019}. In this model, one artificially turns off the tunneling terms between atoms of the same sublattice in the two graphene sheets ($AA, BB$). The resulting Hamiltonian has an additional chiral symmetry (acting as $\s_z$ in sublattice space). One then finds a series of "magic" twist angles \cite{Tarnopolsky2019, ren2021wkb} at which there are eight exactly flat bands (two per spin and graphene valley index) at zero energy, which could be chosen to have a definite Chern number $C=\pm 1$ and sublattice polarization. The Chern/sublattice polarization, therefore, serves as a good quantum number for the flat bands (a review of the model symmetries with and without the chiral approximation is given in \cite{Bultinck2020,ledwith2021strong}). 

\begin{figure}
    \includegraphics[scale=.6]{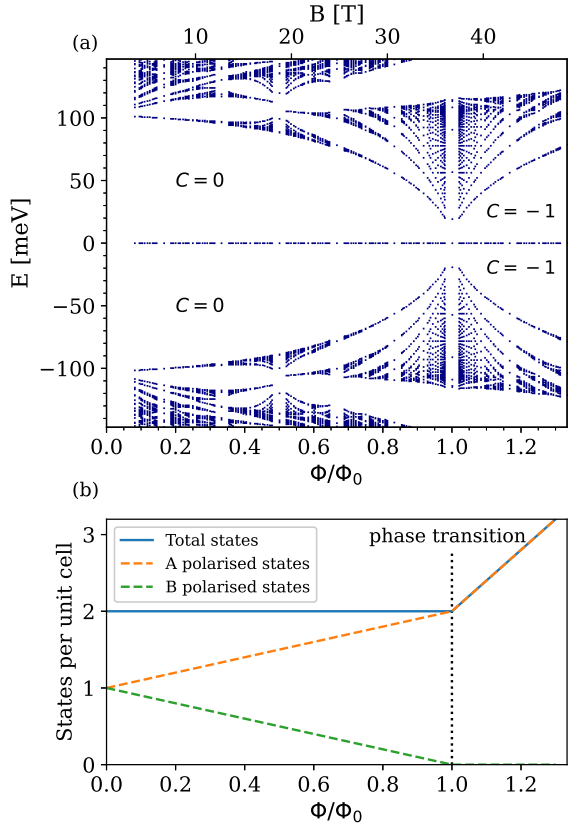}
    \caption{(a) Hofstadter diagram (generated using the method of \cite{Bistritzer2011,Zhang2019}) for the cMATBG model. The bands remain exactly flat for any applied external magnetic field. For $\Phi=\Phi_0$, the gap closes and the total Chern number of the flat bands changes from 0 to 2. (b) Number of states per moir\'e unit cell as a function of magnetic flux. After the phase transition, the number grows as it would for two decoupled Landau levels.}
    \label{fig:hofstadter}
\end{figure}

In this work we use the chiral model at the magic angle to study the interplay between topology, exactly flat bands, and an external perpendicular magnetic field $b$ in twisted bilayer graphene. For the rest of the text we focus on a single valley and ignore the spin. We make here several observations:
\begin{itemize}
    \item The application of a perpendicular field does not make the bands disperse but rather leaves them exactly flat. The magnetic field does not change the Chern number of each band or its association with a particular sublattice. This was first shown by \cite{PhysRevB.103.155150}. Here it is shown by a different method.
    \item The main effect of the magnetic field is to change the number of states within each band.  While at zero field the number of states per unit cell is $n=1$, this number changes to $n(b)=1\pm C\F/\F_0$, where $\F$ is the flux per unit cell and $\F_0$ is the flux quantum. This change contradicts the intuitive expectation that a Bloch band should have one state per unit cell, i.e., $n(b)\equiv1$. It is, however, necessitated by the bands being Chern bands, just as the degeneracy of a Landau level must be linear in the magnetic field. If one imagines filling one such band, and adiabatically turning on a magnetic field $b(t)$ confined to within a large circle in the sample, the azimuthal electric field generated by the time variation of $b$ generates a Hall current into the circle. The in-flowing electrons must have states to occupy, and thus the number of states in the band must vary. Equivalently, this may be seen as an example for the Streda formula, applied separately to each band \cite{Streda1982}:
\begin{equation}
\label{streda}
\s_{x y}=-\left(\frac{\partial n}{\partial b}\right)_\mu,
\end{equation}
where $n$ is the electron density. We absorb the charge $e$ into the definition of the fields ($b=eB$) and assume throughout the work that $\hbar=c=1$. The formula is valid as long as the chemical potential $\mu$ is in the gap and the Chern number remains a good quantum number.
    \item As long as $\abs{\F}<\F_0$ and all bands are full, the total Chern number of the zero-energy bands is zero. As $\F=\F_0$ the $C=-1$ band is emptied, and the Hall conductivity changes to a non-zero integer. Thus, at $\F=\F_0$, the gap must close to allow for this change. Due to the chiral symmetry, the gap closes from both sides of the $E=0$ line (see Fig. \ref{fig:hofstadter}), and we obtain a single Dirac cone in the $\F=\F_0$ band structure. 
    \item We derive exact wave functions for the exactly flat Chern bands with the applied magnetic field. They are given in general in the form
    \begin{equation}
    \label{LL_similarity}
        \y(\bm r)=f(z) e^{-\frac{b_{\rm eff}}{4}\qty|z|^2} G(\bm r), 
    \end{equation}
    where $z=x+iy$ and $f(z)$ is analytic. The $b$-independent functions $G(\bm r)$ are spinors in the layer indices \cite{PhysRevResearch.3.023155}. They are invariant under translations up to a position-dependent phase and inherit the $C_3$ symmetry of the original model. The effective magnetic field $b_{\rm eff}$ is 
    \begin{equation}
    \label{b_eff}
    b_{\rm{eff}}=b+C\frac{2\p}{A_M},
    \end{equation}
    where $C$ is the Chern number of the layer and $A_M$ is the moir\'e unit cell area. The restricted effect of $b$ on the wave function may be indicated by a direct substitution of (\ref{LL_similarity}) in the Schrödinger equation that corresponds to the Hamiltonian (\ref{H_chiral}). This formulation makes explicit the relation between the chiral model wavefunctions and the lowest Landau level (LLL), in which $G(\bm r)\equiv1$, and generalizes the results of \cite{Tarnopolsky2019,wang2021exact} to a general magnetic field. The form given in \eqref{LL_similarity} describes the wavefunctions in the $C=+1$ bands. Different bands are then obtained by $C_2$ and $C_2\TT$ symmetries. \\
    \end{itemize}

Following \cite{Tarnopolsky2019}, we write the chiral MATBG (cMATBG) Hamiltonian in the presence of a magnetic field as
\begin{equation}
\label{H_chiral}
\begin{aligned} \HH_\mathrm{chiral} & =\begin{pmatrix}0 & \DD_{-b}^{*}\left(-\bm{r}\right)\\
\DD_b\left(\bm{r}\right) & 0
\end{pmatrix},\\
\DD_b\left(\bm{r}\right) & =
\begin{pmatrix}
-2ik_\q^{-1}\qty(\bar{\partial}+ z b/4) & \a U\left(\bm{r}\right)\\
\a U\left(-\bm{r}\right) & -2ik_\q^{-1}\qty(\bar{\partial}+ z b/4)
\end{pmatrix},
\end{aligned}
\end{equation}
where $\bar{\partial}=\frac{1}{2}(\partial_x+i\partial_y)$ and $U(\bm r)$ is a position dependent tunneling potential \footnote{See Supplemental Material at [URL will be inserted by publisher] for a derivation of $\HH_\mathrm{chiral}$}. The Hamiltonian acts on the bispinor $\qty(\y(\bm r),\chi (\bm r))$ where each spinor lives on a different graphene sublattice ($A/B$) and the "spins" are the layer indices. We use the symmetric gauge $\bm A=\frac{b}{2}\qty(y\bm{\hat{x}}-x\bm{\hat{y}})$ (so that $b>0$ implies a magnetic field in the $-\bm{\hat{z}}$ direction for negatively-charged carriers), and assume the commensurability condition $b=\frac{p}{q}\frac{2\p}{A_M}$, where $p,q$ are coprime integers. This assumption allows us to label the wavefunctions by common eigenvalues (lattice momenta) of  $\tilde{T}_1^q,\tilde{T}_2$, where $\Tilde T_i$ are the magnetic translation operators by
\begin{equation}
\label{translations}
\begin{aligned}
\bm{a}_{1} & =\frac{4\p}{3k_{\q}}\left(0,-1\right),\\
\bm{a}_{2} & =\frac{4\p}{3k_{\q}}\left(\frac{\sqrt{3}}{2},\frac{1}{2}\right).
\end{aligned}
\end{equation}
We provide an explicit form of $\tilde{T}_i$ in the Supplementary Material (SM) \cite{SM}. In this form of $\HH_{\rm{chiral}}$ the translations act as nonsymmorphic symmetries where a phase difference $\bar{U}=(1,e^{2\pi i/3})$ must be added between the layers. This phase can be removed by gauge choice and will not be considered further.\\
In the absence of a magnetic field, the authors of \cite{Tarnopolsky2019} found that for specific values of $\alpha$, the "magic angles" of the system, the model has two exactly flat bands at zero energy. The bands can be labeled by the sublattice index and have a definite Chern number $C=1$ for the $A$ sublattice and $C=-1$ for the $B$ sublattice. The flatness is sensitive to model parameters, and either changing $\a$ away the magic value or adding any $AA$ tunneling will result in a finite bandwidth. One would then expect that any addition of a magnetic field $b$ will also destroy the exact flatness of the bands. As we now show, this is not the case. 

\paragraph*{Landau level correspondence and exact wavefunctions.\hbox{---}}
Let us consider the relationship between the zero-energy wavefunctions and the wavefunctions of the lowest Landau level (LLL), which are the ground states of the Landau level Hamiltonian
\begin{equation}
    H_{\rm LL} = \qty(-i\bm \na +\bm A)^2.
\end{equation}
We recall that in the symmetric gauge the LLL wave function  are given by Eq. (\ref{LL_similarity}) with $G({\bf r})=1$ and $b_{eff}=b$. 
For cMATBG in the absence of an external magnetic field, one can write an exact correspondence between any wavefunction in the perfectly flat Chern band and a function in the LLL at $b=\frac{2\p}{A_M}$: As was shown by \cite{Tarnopolsky2019}, at the magic angle the wavefunction $\y_K(\bm r)$, corresponding to the $A$-sublattice polarized wavefunction at the $K$ point, has a zero at $\bm r_0=(2\bm a_1+\bm a_2)/3$. Since the kinetic term of $\DD_0(\bm r)$ is proportional to the identity matrix we can use $\y_K(\bm r)$ to construct zero-modes of $\DD_0(\bm r)$ of the form \cite{Tarnopolsky2019,Ledwith2020}
\begin{equation}
\label{cMATBG_wavefunctions}
    \y(\bm r) = \tilde{f}(z)\frac{\y_K(\bm r)}{\vartheta_1\qty(\frac{z-z_0}{a_1} \mid \w)},
\end{equation}
where $\w=e^{2\p i/3},z_0 = r_{0,x}+i r_{0,y}$ (we use non-bold symbols for the complex versions of the vectors $a_i=\bm a_{i,x}+ i \bm a_{i,y}$), and $\vartheta_1(z\mid\tau)$ is the Jacobi theta function. The definition of $\vartheta_1(z\mid\tau)$ is provided in the SM \cite{SM}. Here we note that 
\begin{equation}
\begin{aligned}
    \vartheta_1(z\mid\tau)&=0 & & \leftrightarrow & z & = m+n\tau
\end{aligned}
\end{equation}
for integer $m,n$. In this way the zeros of the theta function cancel the zeros of $\y_K(\bm r)$, making the wave functions (\ref{cMATBG_wavefunctions}) normalizable. Furthermore, the function $\vartheta_1(z\mid\tau)$ is periodic in $\Re(z)$, with a period of one. When $\Im(z)$ is taken to infinity, the function oscillates, with an envelope that diverges as $e^{\frac{\p(\Im z)^2}{\tau}}$. This envelope makes (\ref{cMATBG_wavefunctions}) decay as a Gaussian in the $y$-direction. Writing ${\tilde f}=fe^{\frac{\pi}{2 A_M}(z-2z_0)^2}$ we can write the wave function as a product of an analytic function $f(z)$ multiplied by a function that decays like a Gaussian in all directions, similar to the wave function of the lowest Landau level in the symmetric gauge \footnote{The factor of 2 before $z_0$ assures that the decay is like $e^{\frac{\pi}{2 A_M}\abs{z}^2}$ and not $e^{\frac{\pi}{2 A_M}\abs{z-z_0}^2}$}. With this, we find that there is a correspondence between LLL wavefunctions and wavefucntions of the cMATBG flat bands, given by 
\begin{equation}
\label{zero_field_correspondence}
    f(z)e^{-\frac{2\p}{A_M}\frac{1}{4}\qty|z|^2} \mapsto f(z)\frac{e^{\frac{\pi}{2 A_M}(z-2z_0)^2} \y_K(\bm r)}{\vartheta_1\qty(\frac{z-z_0}{a_1} \mid \w)}.
\end{equation}
The critical property of \eqref{zero_field_correspondence} is that if $f(z)$ is chosen such that the wavefunction on the LHS is an eigenfunction of the magnetic translations, the wavefunctions on the RHS will be eigenfunctions of the translation operators in cMATBG. It therefore maps a set of orthogonal eigenfunctions to a set of orthogonal eigenfunctions and thus provides a basis of wavefunctions for the cMATBG flat bands. 

For $b\neq 0$ one sees that the for any $\y(\bm r)$ of the form \eqref{cMATBG_wavefunctions} the function $e^{-\frac{b}{4}\abs{z}^2}\y(\bm r)$ solves $\mathcal{D}_b\y(\bm r)=0$. The more difficult part of our argument is to obtain the number of orthogonal functions of this form. To do so we extend the correspondence \eqref{zero_field_correspondence} by using the effective magnetic field $b_{\rm eff}$ defined in \eqref{b_eff}, and mapping
\begin{equation}
\label{general_correspondence}
    f(z) e^{-\frac{b_{\rm{eff}}}{4}\qty|z|^2} \mapsto f(z)\frac{e^{\frac{1}{4}\qty(-b\qty|z|^2+\frac{2\p}{A_M} (z-2z_0)^2)}\y_K(\bm r)}{\vartheta_1\qty(\frac{z-z_0}{a_1} \mid \w)}.
\end{equation}
The fact that the RHS gives a zero-mode of $\mathcal{H}_{\rm chiral}$ can be shown by direct substitution. In contrast with \eqref{zero_field_correspondence}, the transformation \eqref{general_correspondence} is not guaranteed to preserve orthogonality between the wavefunctions: we see that in the former case both sides can be labels by eigenvalues of the translation/magnetic translation operators by $\qty(\bm a_1,\bm a_2)$. In the latter case, on the other hand,  the LLL wavefunctions can be classified by the magnetic translation symmetries $\qty(\frac{q}{p+q}\bm a_1,\bm a_2)$ while the cMATBG model has only the magnetic translation symmetries $\qty(q\,\bm a_1,\bm a_2)$. This means that the correspondence gives $p+q$ wavefunctions for each $\rm k$ in the magnetic moir\'e Brillouin zone of TBG, whose orthogonality is not constrained by symmetries. Nevertheless, the resulting $p+q$ wavefunctions remain linearly independent. Otherwise, we could construct a finite linear combination of them that will be mapped to zero under \eqref{general_correspondence}, which is clearly impossible for any non-zero $f$. This shows that the Landau level correspondence gives us $p+q$ states per $\bm k$ in the cMATBG magnetic moir\'e Brillouin zone, or equivalently $1+\frac{\F}{\F_0}$ per unit cell. We can identify $G(\bm r)$ as presented in \eqref{LL_similarity} as 
\begin{equation}
    G(\bm r)=\frac{e^{\frac{\p}{2A_M}\qty(\qty|z|^2+ (z-2z_0)^2)}\y_K(\bm r)}{\vartheta_1\qty(\frac{z-z_0}{a_1} \mid \w)}.
\end{equation}
We can also explicitly write the Bloch wavefunctions obtained from the above correspondence as
\begin{equation}
\label{magnetic_wavefunctions}
\begin{aligned}
    \y_{\bm k ,n}(\bm r) &= \vartheta_1\qty(\frac{q+p}{q}\frac{z-z_0}{a_1}+\frac{k}{b_2} + \frac{n\w}{q}+\frac{i a_1 a_2 b}{8\p}\mid \frac{q+p}{q}\omega) \\
    &\times e^{\qty(k_y+2\pi i \frac{n-p/2}{q a_1}) z} \cdot\frac{e^{-\frac{b}{4}\qty(\qty|z|^2+z^2)}\y_K(\bm r)}{\vartheta_1\qty(\frac{z-z_0}{a_1} \mid \w)},
\end{aligned}
\end{equation}
where $n=1,...,p+q$ and $\rm b_2=k_\q\qty(\sqrt{3},0)$ is a reciprocal lattice vector. The functions $\y_{\bm k ,n}$ for different $n$ are not orthogonal but form a linearly independent set. They can be made orthogonal by a Gram-Schmidt process. \\
We note that the above analysis works for negative and positive $b$ alike, and that the result for $B$ sublattice can be obtained by applying the $C_2 \TT$ symmetry together with $b\to-b$. The number of $B$-polarized states will then be $1-\frac{\F}{\F_0}$. For both cases, this agrees with the arguments given above.\\
For $\F>\F_0$ we find that there are additional $p-q$ $A$-polarized zero-modes so that the total number of zero-modes is $2\frac{\Phi}{\Phi_0}$ per unit cell (as we would have for two uncoupled Landau levels). These additional functions cannot, however, be obtained from \eqref{general_correspondence} and we need to seek a more general form of zero-energy wavefunctions for \eqref{H_chiral}. This form is discussed in the SM \cite{SM}. By the Streda formula \eqref{streda}, we can conclude that the total Chern number of the zero-energy wavefunctions becomes $C=2$ for $\F>\F_0$. The transition of the Chern number should be accompanied by a closure of the gap, which results in two additional zero-modes at the $\Gamma$ point $\bm k_0 =(0,-k_\q)$. In the SM we show how one can write explicit forms for the additional zero-modes \cite{SM}.\\
Notice also that off the magic angle we have exactly $2\frac{\F}{\F_0}$ zero-energy wavefunctions, which are all $A$-sublattice polarized, and are given by
\begin{equation}
    \y_{z.m.}(\bm r)=\nu(\bm r)\y_{K/K'}(\bm r),
\end{equation}
where $\nu(\bm r)$ is a function in the LLL (Eq. (\ref{LL_similarity}) with magnetic field $b$ and $G({\bf r})=1$) and $\y_{K/K'}(\bm r)$ are the zero-energy $A$-polarized wavefunctions at zero magnetic field, at the $K$ and $K'$ points of the moir\'e Brillouin zone.
\paragraph*{Relation to the Atiyah-Singer index theorem.\hbox{---}}
An elegant way to reproduce the number of zero energy wavefunctions is by means the Atiyah-Singer index theorem \cite{Atiyah1984,Nakahara1990}. As was pointed out in \cite{PhysRevLett.108.216802}, the Hamiltonian \eqref{H_chiral} is a Hamiltonian of a Dirac electron in a $U(1)\times S U(2)$ gauge potential, where the tunneling term couples to the $S U(2)$ sector and the external field couples to the $U(1)$ sector. The non-abelian gauge potential $\bm{\mathcal{A}}$ is given by
\begin{equation}
\begin{aligned}
    \bm{\mathcal{A}}&= \bm A I + \bm A_{\rm n.a.}^i \s_i, \\
    \bm A_{\rm n.a.}^i&= \frac{1}{2}\qty(\Re[\Tr(\s_i \mathcal{U})],\Im[\Tr(\s_i \mathcal{U})]) ,\\
    \mathcal{U}&=i\a k_\q
    \begin{pmatrix}
        0 & U(\bm r) \\
        U(-\bm r) & 0
    \end{pmatrix},
\end{aligned}
\end{equation}
where $I$ is the identity matrix and summation over the three Pauli matrices is implied in the first line. At zero energy, eigenstates reside purely on one sublattice, and we can discuss the number $n_{A/B}$ of zero-energy states on the two sublattices (per unit cell). Assuming periodic boundary condition, the difference $n_A-n_B$ between the number of zero-energy modes (per unit cell) in the $A$ and $B$ sublattices is a form of a Chiral anomaly. This difference can be obtained exactly by means of the Atiyah-Singer index theorem \footnote{We follow the form for twisted spin bundles, provided in Eq. (12.86) of \cite{Nakahara1990}}:
\begin{equation}
    n_A-n_B=\frac{1}{2\p N_s}\int\Tr \mathcal{F}_{x y}d\bm r,
\end{equation}
where $N_s$ is the number of unit cells and $\mathcal{F}_{x y}$ is the curvature associated with $\bm{\mathcal{A}}$, given by
\begin{equation}
    \mathcal{F}_{x  y}=\partial_x\mathcal{A}_y-\partial_y\mathcal{A}_x+\frac{1}{2}[\mathcal{A}_x,\mathcal{A}_y].
\end{equation}
The non-abelian part has no contribution to the trace \footnote{The vanishing contribution of the $SU(2)$ part is related to the fact that $\Pi_2(SU(2))=0$.}, while the integral of the abelian part gives
\begin{equation}
    n_A-n_B=2\frac{\F}{\F_0}.
\end{equation}
This result does not depend on the system being at the magic angle or on the magnetic field being uniform in space. The total number $n_A+n_B$ of zero-energy eigenstates may vary considerably, being $2\max (\frac{\F}{\F_0},1)$ at the magic angle and $2\frac{\F}{\F_0}$ off the magic angle, but the difference $n_A-n_B$ (in the chiral model) is governed by the topology.


\paragraph*{Fractional Chern Insulators.\hbox{---}}
Since the zero energy bands of the model can be presented as Chern bands, one may expect that at partial filling and in the presence of electron-electron interactions the ground state would be a fractional Chern insulator (FCI) in which both time-reversal and flavor symmetries are spontaneously broken \cite{Bernevig2013, Parameswaran2012, Spanton2018}. A recent experiment \cite{xie2021fractional} has shown that indeed an FCI state can be found at fillings $3<n_e<4$ ($n_e$ refers to the number of electrons per unit cell added, counted from charge neutrality), with non-zero magnetic field. These states seem to be connected to zero-field FCI states \cite{abouelkomsan2020particle, Repellin2020, PhysRevB.101.235312} but require an external magnetic field to be stabilized. The analysis of \cite{xie2021fractional}  suggests that the importance of the external magnetic field is that it flattens the Berry curvature, thus making the FCI more favorable \cite{Jackson2015,wang2021exact}. \\
Our expressions for the wave functions of the flat bands in the presence of a perpendicular magnetic field naturally suggest a few observations regarding the FCI states. Starting with the case of one partially filled Chern band, and defining $n$ as the number of electrons per unit cell in that band, the filling fraction of the band is $n/(1+C_p\frac{\F}{\F_0})$, with $C_p$ its Chern number. An FCI of the Jain series \cite{jain1989composite} would form when $n=(1+C_p\frac{\F}{\F_0})\frac{j}{2mj+1}$ with $m,j$ integers. When the partially filled band co-exists with $N_b$ full bands of Chern numbers $C_i$ (with $i=1,...,N_b$), the relation between the total number of electrons per unit cell $n_e$ and the flux density $\frac{\F}{\F_0}$ is 
\begin{equation}
    n_e=\sum_{i=1}^{N_b}(1+C_i\frac{\F}{\F_0})+(1+C_p\frac{\F}{\F_0})\frac{j}{2mj+1},
    \label{Jainseries}
\end{equation}
as expected from the Streda formula \eqref{streda}, for constant $m,j$ the number of electrons varies when the flux is changed, and the slope of the variation is the corresponding Hall conductivity of the incompressible state. 

For $p=1$, for example, we get a series analogous to the Laughlin series, where the analog to the Laughlin wave function (in the chiral limit) is 
\begin{equation}
    \Psi \qty(\{z_i\})=
    \prod_{i<j}(z_i-z_j)^m\prod_i e^{-\frac{b_{\rm eff}}{4}\qty|z_i|^2} G(\bm r_i),
\end{equation}
generalizing the results of \cite{Ledwith2020} to an arbitrary magnetic field. \\
When the FCI forms in a band whose Chern number is negative, $C=-1$, the effective magnetic field \textit{decreases} in magnitude when the external magnetic field is increased. This decrease has an interesting implication on the characteristic 
interaction energy scale of the FCI (with Coulomb interactions), which is 
\begin{equation}
    V=\frac{e^2}{\varepsilon \ell_{B,\rm eff}},
    \label{interaction_scale}
\end{equation}
where, $\ell_{B,\rm eff}^{-2}=b_{\rm eff}$, and $\varepsilon$ is the dielectric constant. 
We expect the energy gap of the FCI to be determined by $V$, up to a proportionality factor that depends on the band-structure geometry \cite{Parameswaran2012,Jackson2015}, but has only weak magnetic field dependence. For large $b$, then, Eq. (\ref{interaction_scale}) suggests that the FCI energy gap may decrease with an increasing magnetic field, in contrast to that of fractional quantum Hall states. In that limit, we expect the FCI state to be destabilized. 
\paragraph*{Discussion.\hbox{---}}
As we focused in this letter on the chiral model Hamiltonian, it is worthwhile to discuss which of our conclusions extend beyond it. In the case where a spontaneous symmetry breaking results in a band with a definite Chern number (as is expected, e.g., in $3<\nu<4$ in TBG), one can still use the effective magnetic field approach to discuss fractional states in this regime. Furthermore, an interesting future direction is to use our results as a starting point for analyzing the stability of FCI states as a function of the magnetic field in the Chiral model and beyond. Finally, the correspondence \eqref{zero_field_correspondence} (and therefore \eqref{general_correspondence}) relies on nothing more but the existence of a kinetic term acting as an anti-holomorphic derivative with no dependence on the spinor direction, together with the zero of $\y_K(\bm r)$. It would be interesting to ask whether one can find additional models exhibiting the same correspondence. 
\begin{acknowledgements}
    We thank Amir Yacoby, Yonglong Xie and Andrew Pierece for enlightening discussions. A.S. acknowledges support from the Israeli Science Foundation Quantum Science and Technology grant no. 2074/19, the CRC 183 of the Deutsche Forschungsgemeinschaft. This project has received funding from the European Research Council (ERC) under the European Union’s Horizon 2020 research and innovation programme (grant agreement No. 788715, Project LEGOTOP).
\end{acknowledgements}

\onecolumngrid
\appendix
\section{Chiral model and symmetries}
\subsection{Derivation}
Here we re-derive the cMATBG Hamiltonian \eqref{H_chiral}. Our derivation follows that of \cite{Tarnopolsky2019}. We begin with the Bistrizer-Macdonald (BM) Hamiltonian \cite{Bistritzer2011a,Bistritzer2011}

\begin{align}
H & =\begin{pmatrix}h\left(-\q/2\right) & T\left(\bm{r}\right)\\
T^\dagger\left(\bm{r}\right) & h\left(\q/2\right)
\end{pmatrix},\\
h & =-iv\bm{\s}_{\q/2}\cdot \bm{\na},\\
T\left(\bm{r}\right) & =w\sum_{j}e^{-i\bm{q}_{j}\cdot\bm{r}}T_{j},
\end{align}
where $\bm{\s}_{\q}=e^{-i\q\s_{z}/2}\left(\s_{x},\s_{y}\right)e^{i\q\s_{z}/2}$,
the scattering matrices $T_{i}$ are
\begin{equation}
\begin{aligned}
T_{1} & =\begin{pmatrix}\k & 1\\
1 & \k
\end{pmatrix}, \\
T_{2,3} & =\begin{pmatrix}\k & e^{\mp i\f}\\
e^{\pm i\f} & \k
\end{pmatrix},
\end{aligned}
\end{equation}
with $\f=2\p/3$ and $\bm{q}_{1}=k_{\q}\left(0,-1\right),\bm{q}_{2,3}=k_{\q}\left(\pm\sqrt{3},1\right)/2$.
We have $k_{\q}=2\sin\left(\q/2\right)k_{D}\approx\q k_{D}$ where
$k_{D}=\frac{4\p}{3\sqrt{3}a_{0}}$ and $a_{0}\approx1.4\si{\angstrom}$
is the distance between atoms in graphene. The scale $w\approx 110\si{meV}$ is the energy scale associated with the tunneling between the layers and the factor $0\le\kappa\le1$ determines the ratio between $AA$ and $AB$ tunneling layers of the graphene. Real world MATBG has $\kappa\approx 0.7$ as a result of lattice relaxation. The lattice and dual lattice of the model are spanned by
\begin{equation}
    \begin{aligned}
    \bm a_1&=\frac{4\p}{3k_\q}(0,-1) & \bm b_1&=\sqrt{3}k_\q \qty(\frac{1}{2},-\frac{\sqrt{3}}{2})\\
    \bm a_2&=\frac{4\p}{3k_\q}\qty(\frac{\sqrt{3}}{2},\frac{1}{2}) & \bm b_2 & = k_\q(\sqrt{3},0).
    \end{aligned}
\end{equation}
The chiral model results from setting $\k=0$.  Under this assumption we can remove the $\theta$ dependence in $h(\q/2)$ by a gauge transformation. We further re-scale the Hamiltonian by defining $\HH=E_0 H$ where $E_0=k_\q w$, define the dimensionless parameter $\a = w/k_\q v$ and rearrange the rows so that the Hamiltonian acts on the spinor $(\y_1,\y_2,\chi_1,\chi_2)$. We obtain the chiral Hamiltonian
\begin{equation}
\label{H_chiral_app}
\begin{aligned} \HH_\mathrm{chiral} & =\begin{pmatrix}0 & \DD_{0}^{*}\left(-\bm{r}\right)\\
\DD_0\left(\bm{r}\right) & 0
\end{pmatrix},\\
\DD_0\left(\bm{r}\right) & =
\begin{pmatrix}
-2i k_\q^{-1}\bar{\partial} & \a U\left(\bm{r}\right)\\
\a U\left(-\bm{r}\right) & -2i k_\q^{-1}\bar{\partial}
\end{pmatrix},
\end{aligned}
\end{equation}
where $z=x+i y,\bar{\partial}=\frac{1}{2}(\partial_x+i\partial_y)$ and $U(\bm r)=e^{i\bm q_1\cdot\bm r}+e^{i\f}e^{-i\bm q_2\cdot\bm r}+e^{-i\f}e^{-i\bm q_2\cdot\bm r}$. Finally, coupling the system to an external magnetic field via the Peierls substitution $-i\bm \na\to -i\bm \na+\bm A$, we obtain \eqref{H_chiral}.
\subsection{Symmetries}
\subsubsection{Point symmetries}
Let us discuss the symmetries of the BM Hamiltonian. We define the Pauli matrices $\s_i,\h_i$ in sublattice and layer space, respectively. The point-symmetries acting within the valley are given by
\begin{equation}
    \begin{aligned}
    C_2\TT:& &\s_x K(\bm r\to-\bm r), \\
    C_3:& &e^{-i\frac{2\p}{3} \s_z}(\bm r\to R_3\bm r), \\
    M:& & \h_x\s_x (y\to -y),
    \end{aligned}
\end{equation}
where $K$ is the complex conjugation operator and $R_3$ is the rotation matrix by $2\p/3$. The Chiral model has the additional chiral symmetry
\begin{equation}
\begin{aligned}
    \mathcal{C}:& &\s_z.
\end{aligned}
\end{equation}
For $b\neq 0$ both the $C_2\TT$ and $M$ symmetries are broken, however, the combination
\begin{equation}
    \begin{aligned}
    MC_2\TT:&& \h_x K(x\to-x)
    \end{aligned}
\end{equation}
is preserved.
\subsubsection{Magnetic translations}
In the presence of a magnetic field, the translation symmetries should be replaced by the magnetic translation operators $\tilde{T}_i$. We define $\tilde{\bm A}$ which satisfies $\partial_i A_j=\partial_j\tilde{A}_i$. For the symmetric gauge $\tilde{\bm A}=-\bm A=\frac{b}{2}\qty(x\hat{\bm y}-y\hat{\bm x})$. The magnetic translation symmetries are then given by
\begin{equation}
    \begin{aligned}
    \tilde{T}_i: e^{i\bm a_i\cdot\bm \tilde{A}(\bm r)}(\bm r\to\bm r+\bm a_i).
    \end{aligned}
\end{equation}
The major difference between the magnetic translations and the translation operators is that in general $\qty[\tilde{T}_1,\tilde{T}_2]\neq 0$. Instead, they satisfy
\begin{equation}
    \tilde{T}_1\tilde{T_2}=e^{i\F}\tilde{T_2}\tilde{T}_1,
\end{equation}
where $\F=bA_M$ is the magnetic flux per unit cell. To be able to define a Bloch basis we assume the commensurability condition $\F=2\p \frac{p}{q}$ where $p,q$ are coprime integers. Under this assumption we have $\qty[\tilde{T}_1^q,\tilde{T}_2]=0$ and we can define a Bloch basis of wavefunctions satisfying
\begin{equation}
\begin{aligned}
    \tilde{T}_1^q\y_{\bm k}(\bm r)&=e^{i q\bm k\cdot \bm a_1} \y_{\bm k}(\bm r) \\
    \tilde{T}_2\y_{\bm k}(\bm r)&=e^{i\bm k\cdot \bm a_2} \y_{\bm k}(\bm r),
\end{aligned}
\end{equation}
where $\bm k$ is now defined on the unit cell of the lattice spanned by $(\frac{1}{q}\bm b_1,\bm b_2)$. That is, the magnetic moir\'e Brillouin zone (mmBZ) is $q$ times smaller. Of course, in practice all measurable quantities (e.g., density of states, Chern number in the band gap) should be continuous in $\F$.

\section{The theta function}
\begin{figure}
    \centering
    \includegraphics[scale=.7]{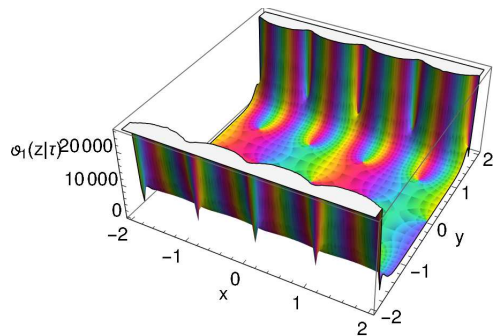}
    \caption{Plot of the function $\vartheta_1(z\mid\omega)$ on the complex plane. The colors signify the phase of the function. One can see that the function is periodic in the $x$ direction, while grows super-exponentially in the $y$ direction, and has zeros for $m+\omega n$.}
    \label{fig:vartheta}
\end{figure}

We give below a few relevant properties of the theta function $\vartheta_1(z\mid\tau)$. It is given by \cite{Whittaker1996}
\begin{equation}
\label{vartheta}
\begin{aligned}
    \vartheta_1(z\mid\tau)&=\sum_{n=-\infty}^\infty (-1)^{n-1/2} e^{i\pi(n+1/2)^2\tau} e^{2\pi i(n+1/2)z} \\
        &=2\sum_{n=0}^\infty(-1)^n e^{i\p(n+1/2)^2 \tau}\sin{(2\p(n+1/2)z)},
\end{aligned}
\end{equation}
where $\tau$ is the half-period ratio and must satisfy $\rm{Im}(\tau)>0$. We plot the function for $\tau=e^{2\p i/3}$ in Fig. \ref{fig:vartheta} The function is analytic in $z$ and satisfies the quasi-periodicity conditions
\begin{equation}
\label{quasiperiodicity}
\begin{aligned}
    \vartheta_1(z+1\mid\tau)&=-\vartheta_1(z\mid\tau), \\
    \vartheta_1(z+\tau\mid\tau)&=-e^{-i\p\tau}e^{-2\p i z}\vartheta_1(z\mid\tau), \\
\end{aligned}
\end{equation}
which can be verified by direct substitution. The form \eqref{vartheta} shows that $\vartheta_1(z\mid\tau)$ must have a zero for $z=0$. By the quasi-periodicity we therefore find that
\begin{equation}
\begin{aligned}
    \vartheta_1(z\mid\tau)&=0 & & \leftrightarrow & z & = m+n\tau.
\end{aligned}
\end{equation}
The quasi-periodicity conditions also tell us that the function is periodic in $\rm{Re}(z)$. By Liouville's theorem it must therefore be unbounded as $\rm{Im}(z)\to\pm\infty$. Indeed, we can define the function
\begin{equation}
    \tilde{\vartheta}(x,y\mid\tau)=e^{-\frac{\pi}{\rm{Im}(\tau)}y^2}\vartheta_1(x+iy\mid\tau),
\end{equation}
for which $\abs{\tilde{\vartheta}(x,y\mid\tau)}$ is periodic on the lattice $n+\tau m$. \\
The exponential growth of $\vartheta_1(z\mid\tau)$ is of major importance to our discussion, as it enables the writing of the cMATBG wavefunctions of the form \eqref{cMATBG_wavefunctions}. Had $1/\vartheta_1(z\mid\tau)$ not decayed at infinity we could not find analytic functions $\tilde{f}(z)$ that would make $\eqref{cMATBG_wavefunctions}$ normalizable. 
\section{Number of zero-energy states}
While we showed in the main manuscript that the functions obtained from \eqref{general_correspondence} are indeed a set of zero-energy wavefunctions of $\DD_b$ we have not shown that these are the only solutions. In fact, we shall see that additional solutions emerge for $\F\ge\F_0$. To see these additional solutions, let us apply the linear transformation
\begin{equation}
\begin{aligned}\y_{\bm{k}}\left(\bm{r}\right) & =\mathcal{S}\left(\bm{r}\right)\h_{\bm{k}}\left(\bm{r}\right),\\
\mathcal{S}\left(\bm{r}\right) & =\begin{pmatrix}-\y_{K,2}^{*}\left(\bm{r}\right) & \y_{K,1}\left(\bm{r}\right)\\
\y_{K,1}^{*}\left(\bm{r}\right) & \y_{K,2}\left(\bm{r}\right)
\end{pmatrix},
\end{aligned}
\end{equation}
where the indices are layer indices. The requirement that $\y_{\bm k}(\bm r)=\nu(\bm r)\y_K(\bm r)$ for some scalar function $\nu(\bm r)$ is equivalent to $\eta_{\bm k,1}=0$. For $\y_{\bm{k}}$ to be a zero-mode $\h_{\bm{k}}$ must satisfy
\begin{equation}
\begin{pmatrix}
\label{eq:eta_hamiltonian}
-2i\qty(\bar{\partial}-\frac{b z}{4}+\left(\bar{\partial}\log\r_{K}\right)) & 0 \\
h\left(\bm{r}\right) & -2i\qty(\bar{\partial}-\frac{b z}{4})
\end{pmatrix}\h_{\bm{k}}\left(\bm{r}\right)=0,
\end{equation}
where $\r_{K}\left(\bm{r}\right)=\y^\da_{K}\left(\bm{r}\right)\y_{K}\left(\bm{r}\right)$ is the local density of $\y_K(\bm r)$ and the function $h(\bm r)$ is 
\begin{equation}
\begin{aligned}
h(\bm r)&=\r^{-1}_K(\bm r)(\y_{K,2}^*)^2(2i\bar{\partial}(\y_{K,1}^*/\y_{K,2}^*) +\a(U(-\bm r)-U(\bm r)(\y_{K,1}^*/\y_{K,2}^*)^2.
\end{aligned}
\end{equation}
For $\Phi < \Phi_{0}$ (or equivalently $p<q$) we find that the only possible solution for $\eta_{\bm k,1}$ is $\h_{\bm{k},1}\left(\bm{r}\right)\equiv0$.
This can be shown by writing $\eta_{\bm k,1}$ in a general form compatible with \eqref{eq:eta_hamiltonian} as
\begin{equation}
\label{eta1_solution}
    \h_{\bm{k},1}(\bm{r})=g_{\bm k}(z)\frac{\vartheta_1\qty(\frac{z-z_0}{a_1}\mid\tau)e^{-\frac{1}{4}\qty(b\abs{z}^{2}-\frac{2\p}{A_M}(z-2z_0)^2)}}{\r_{K}\left(\bm{r}\right)}
\end{equation}
for an analytic function $g_{\bm k}(z)$. In order for $\y_{\bm k}$ to be normalizable $g_{\bm k}(z)$ must have no poles. For $\eta_{\bm k,1}(\bm r)$ to be finite at infinity $g_{\bm k}(z)$ must also decay at infinity as $e^{\frac{1}{4}\qty(\frac{2\p}{A_M}-b)\abs{z}^2}$. For $\F<\F_0$ the only normalizable solution is then $\h_{\bm{k},1}=0$. On the other hand, for $p>q$ we get by a similar correspondence to \eqref{general_correspondence} that there are $\frac{\F}{\F_0}-1$ linearly independent solutions for $\h_{\bm k,1}$ which give additional wavefunctions that are not of the form $\nu(\bm r)\y_K(\bm r)$ and are therefore not captured by \eqref{magnetic_wavefunctions}. All in all, we find that for $\F>\F_0$ there are $2\frac{\F}{\F_0}$ zero-energy wavefunctions per unit cell that are all $A$-sublattice polarized. 
\section{Gap closing}
Here we discuss the gap-closing at $\F=\F_0$ and find the additional wavefunctions associated with it. For the gap to close we need a Dirac cone to form around a specific point in the magnetic Brillouin zone, which requires that the equation $\HH_{\rm{chiral}}$ has two additional zero-modes for a specific $\bm k$. Indeed, while for general $\bm{k}$ we have no $B$-polarized states, and two $A$-polarized states given by \eqref{magnetic_wavefunctions}, we find that at the $\Gamma$ point $\bm k_0 =(0,-k_\q)$ (in our gauge choice) there are two additional wavefunctions, one on each sublattice. \\
We start with the $B$ polarized function, which can be obtained by calculating the $A$-polarized wavefunction at $\F=-\F_0$ and noting that the Hamiltonian is invariant under the combination of $C_2\TT$ with $b\to-b$. From \eqref{general_correspondence} we see that for $b_{\rm{eff}}=0$ the only normalizable wavefunction to be obtained from the correspondence is the one with $f(z)\equiv 1$. This constrains the $B$-polarized zero-mode to
\begin{equation}
    \chi_{\bm k_0}(-\bm r)^* = 
    \frac{e^{\frac{b}{4}\qty(\qty|z|^2+(z-2z_0)^2)}\y_K(\bm r)}{\vartheta_1\qty(\frac{z-z_0}{a_1} \mid \w)}.
\end{equation}
The $A$-polarized wavefunction cannot be obtained from \eqref{general_correspondence}, and is instead obtained from the solutions of \eqref{eq:eta_hamiltonian} with non-zero $\eta_{\bm k,1}$. In the form \eqref{eta1_solution} we see similarly to the $B$-polarized case that the only solution at $b=\frac{2\p}{A_M}$ is obtained from $g_{\bm{k}}=1$. This gives again the lattice momentum $\bm k_0$, as expected.
\twocolumngrid
\bibliography{MagcTBG}
\end{document}